\documentclass[aps,prd,twocolumn,groupedaddress,showpacs]{revtex4}
\usepackage{epsf,epsfig,graphicx}
\def\edth{\;\raise1.0pt\hbox{$'$}\hskip-6pt\partial\;}
\def\baredth{\;\overline{\raise1.0pt\hbox{$'$}\hskip-6pt
\partial}\;}
\def\gsim{~\rlap{$>$}{\lower 1.0ex\hbox{$\sim$}}}

\newcommand{\be}{\begin{equation}}
\newcommand{\ba}{\begin{eqnarray}}
\newcommand{\ee}{\end{equation}}
\newcommand{\ea}{\end{eqnarray}}

\newcommand{\fr}{\frac}
\newcommand{\tF}{\tilde{F}}

\begin{document}

\title{Cosmic Birefringence Fluctuations and Cosmic Microwave Background $B$-mode Polarization}

\author{Seokcheon Lee$^{1}$, Guo-Chin Liu$^{2}$, and Kin-Wang Ng$^{3,4}$}

\affiliation{
$^1$Korea Institute for Advanced Study, Seoul 130-722, Korea\\
$^2$Department of Physics, Tamkang University, Tamsui, New Taipei City 25137, Taiwan\\
$^3$Institute of Physics, Academia Sinica, Taipei 11529, Taiwan\\
$^4$Institute of Astronomy and Astrophysics, Academia Sinica, Taipei 11529, Taiwan
}

\vspace*{0.6 cm}
\date{\today}
\vspace*{1.2 cm}

\begin{abstract}
Recently, BICEP2 measurements of the cosmic microwave background (CMB) $B$-mode polarization has indicated the presence of primordial gravitational waves at degree angular scales, inferring the tensor-to-scalar ratio of $r=0.2$ and a running scalar spectral index, provided that dust contamination is low. In this {\em Letter}, we show that the existence of the fluctuations of cosmological birefringence can give rise to CMB $B$-mode polarization that fits BICEP2 data with $r<0.11$ and no running of the scalar spectral index. When dust contribution is taken into account, we derive an upper limit on the cosmological birefringence, $A\beta^2<0.0075$, where $A$ is the amplitude of birefringence fluctuations that couple to electromagnetism with a coupling strength $\beta$.

\end{abstract}

\pacs{98.70.Vc, 98.80.Es}
\maketitle

The spatial flatness and homogeneity of the present
Universe strongly suggest that a period of de Sitter expansion
or inflation had occurred in the early Universe~\cite{olive}.
During inflation, quantum fluctuations of the inflaton
field may give rise to energy density perturbations (scalar
modes)~\cite{pi}, which can serve as the seeds for the formation
of large-scale structures of the Universe. In addition,
a spectrum of gravitational waves (tensor modes) is produced
from the de Sitter vacuum~\cite{star}.

Gravitational waves are very weakly coupled to matter,
so once produced, they remain as a stochastic background
until today, and thus provide a potentially important probe
of the inflationary epoch. Detection of these primordial
waves by using terrestrial wave detectors or the timing
of millisecond pulsars~\cite{krau} would indeed require an experimental
sensitivity of several orders of magnitude beyond
the current reach. However,
horizon-sized tensor perturbation induces large-scale temperature
anisotropy of the cosmic microwave background
(CMB) via the Sachs-Wolfe effect~\cite{sach}. In addition, the tensor modes uniquely induce CMB $B$-mode
polarization that is the primary goal of ongoing and future
CMB experiments~\cite{weiss}.

Recently, WMAP+SPT CMB data has placed an upper limit on
the contribution of tensor modes to the CMB anisotropy, in
terms of the tensor-to-scalar ratio, which is $r<0.18$ at $95\%$ confidence level, tightening to $r<0.11$
when also including measurements of the Hubble constant
and baryon acoustic oscillations (BAO)~\cite{story}. Planck
Collaboration XVI has quoted $r<0.11$ using a combination
of {\em Planck}, SPT, and ACT anisotropy data, plus WMAP polarization; however, the constraint
relaxes to $r<0.26$ ($95\%$ confidence) when running of the scalar spectral index
is allowed with $dn_s/d \ln k = -0.022\pm 0.010$ ($68\%$)~\cite{planck}.
More recently, BICEP2 CMB experiment has found an excess of $B$-mode power at degree angular scales,
indicating the presence of tensor modes with $r=0.20^{+0.07}_{-0.05}$ and $d n_{s}/ d \ln k=-0.028\pm0.009$~\cite{bicep2}.
If this result is confirmed, it would give very strong support to inflation model and open a new window for probing the inflationary dynamics.
However, lately a joint analysis of BICEP2/Keck Array and {\em Planck} data has indicated that the BICEP2 $B$-mode signal could be mainly due to dust foreground~\cite{b+p}. 

In this {\em Letter}, we investigate another source for generating CMB $B$-mode polarization. The generated $B$-mode power spectrum can explain the BICEP2 excess $B$-mode power, while complying to the limit $r<0.11$ and $dn_s/d {\rm ln} k = 0$.
Here we consider a nearly massless pseudoscalar $\Phi\equiv M\phi$ that couples to the electromagnetic field strength via
$(-\beta/4)\phi F_{\mu\nu} \tF^{\mu \nu}$, where $\beta$ is a coupling
constant and $M$ is the reduced Planck mass. The effect of this coupling to CMB polarization has been previously studied in different contexts such as new high-energy physics~\cite{pospelov}, a massless pseudo-Nambu-Goldstone spectator field~\cite{Caldwell}, and scalar quantum fluctuations of the vacuum-like cosmological constant in an axiverse~\cite{LLN2,Zhao}. It is well known that the above $\phi$-photon interaction may lead to
cosmic birefringence~\cite{carroll} that induces rotation of the polarization plane of the CMB,
thus resulting in a conversion of $E$-mode into $B$-mode polarization without affecting the temperature anisotropies~\cite{lue,LLN}.
For such a pseudoscalar, we consider the contribution of $\phi$ perturbation to cosmic birefringence,
resulted from dressed photon propagators with $\phi$ perturbation. Furthermore, $\phi$-photon elastic scatterings can lead to
CMB lensing; however, this is a second-order effect which is subdominant to the gravitational lensing of large-scale structures in the present consideration.

We assume a conformally flat metric,
$ds^2=a^2(\eta) (d\eta^2- d \vec{x}^2)$,
where $a(\eta)$ is the cosmic scale factor and $\eta$ is the conformal time
defined by $dt=a(\eta)d\eta$.
The $\phi F \tF$ term leads to a rotational velocity of the polarization plane of a photon
propagating in the direction $\hat{n}$~\cite{carroll},
\be
\omega(\eta, \vec{x})=-{\beta\over2}\left(\fr{\partial \phi}{\partial\eta}+
\vec{\nabla} \phi \cdot \hat{n}\right).
\label{rot}
\ee
Thomson scatterings of anisotropic CMB photons by free electrons
give rise to linear polarization, which can be described
by the Stokes parameters $Q(\eta, \vec{x})$ and $U(\eta, \vec{x})$.
The time evolution of the linear polarization is governed by the collisional Boltzmann equation,
which would be modified due to the rotational velocity of the polarization plane~(\ref{rot}) by including
a temporal rate of change of the Stokes parameters:
\be
\dot Q\pm i \dot U = \mp i2\omega \left( Q\pm i U \right),
\label{QUeq}
\ee
where the dot denotes $d/d\eta$.
This gives a convolution of the Fourier modes of the Stokes parameters
with the spectral rotation that can be easily incorporated into the Boltzmann code.

Now we consider the time evolution of $\phi$.
Decompose $\phi$ into the vacuum expectation value and the perturbation:
$\phi(\eta,\vec{x}) =\bar{\phi}(\eta) + \delta \phi (\eta, \vec{x})$.
For the metric perturbation, we adopt the synchronous gauge:
$ds^2= a^2(\eta) \{d\eta^2- [\delta_{ij}+h_{ij}(\eta, {\vec x})]dx^i dx^j\}$.
Neglecting the back reaction of the interaction, we obtain the mean field evolution as
\be
\ddot{\bar\phi} + 2 {\cal H} \dot{\bar\phi} +
\fr{a^2}{M^2} \fr{\partial V}{\partial \bar\phi}= 0\, ,\label{phi0eq}
\ee
where ${\cal H} \equiv \dot{a} / a$ and $V(\phi)$ is the scalar potential.
The equation of motion for the Fourier mode $\delta\phi_{\vec k}$ is given by
\be
\ddot{\delta\phi}_{\vec k}
+ 2 {\cal H} \dot{\delta\phi}_{\vec k} + \left( k^2+
\fr{a^2}{M^2} \fr{\partial^2 V}{\partial \bar\phi^2}\right) {\delta\phi}_{\vec k}=
-{1\over2} {\dot h_{\vec k}} {\dot{\bar\phi}}\, .
\label{fourierphi}
\ee
where $h_{\vec k}$ is the Fourier transform of the trace of $h_{ij}$.

\begin{figure}[htbp]
\centerline{\psfig{file=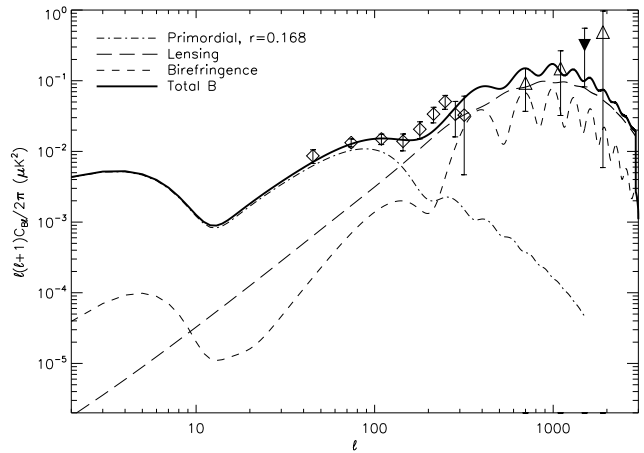, width=9cm}
} \caption{
Cosmological birefringence induced $B$-mode power spectrum through the
perturbed nearly massless scalar field with $A\beta^2=0.0046$ (short-dashed).
Also shown are the theoretical power spectra of lensing induced
$B$ modes (long-dashed) and gravity-wave induced $B$ modes (dot-dashed) with $r=0.168$. The thick solid curve is the best-fitting
averaged $B$-mode band powers that are the sum of these three $B$-mode power spectra convolved with the BICEP2 ($l<400$) and the POLARBEAR ($l>400$) window functions. BICEP2 data~\cite{bicep2} (diamonds) and POLARBEAR data~\cite{polarbearB} (triangles and an inverted solid triangle representing the absolute value of a negative band) are shown.}
\label{fig:BMODE}
\end{figure}

\begin{figure}[htbp]
\centerline{\psfig{file=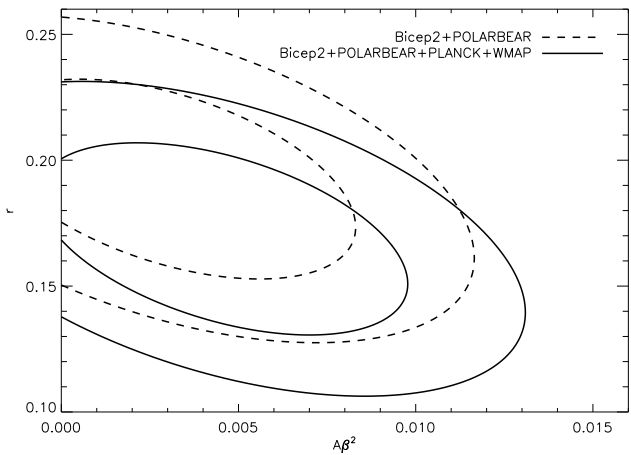, width=9cm}
} \caption{Likelihood plot of the parameters $r$ and $A\beta^2$, showing $1$-sigma and $2$-sigma contours. Solid contours use WMAP, {\em Planck}, BICEP2, and POLARBEAR data, with the maximum likelihood values of $r=0.168$ and $A\beta^2=0.0046$. Dashed contours use BICEP2 and POLARBEAR data only for comparison.}
\label{likelihood}
\end{figure}

\begin{figure}[htbp]
\centerline{\psfig{file=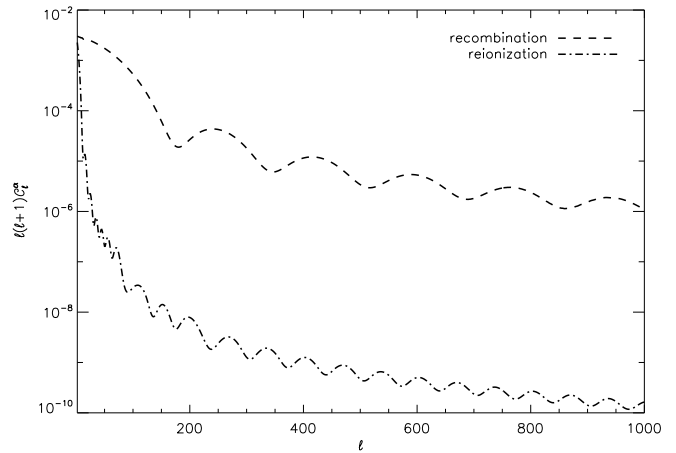, width=9cm}
} \caption{Rotation power spectra at the recombination and the reionization with $A\beta^2=0.0046$.}
\label{fig:alpha}
\end{figure}

If $\phi$ is nearly massless or its effective mass is less than the present Hubble parameter,
the mass term and the source term in Eq.~(\ref{fourierphi}) can be neglected.
In this case, $V(\phi)$ can be either null or behaves just like a cosmological constant
with $\dot{\bar\phi}=0$. However, its perturbation is dispersive and can be cast into
${\delta\phi}_{\vec k}(\eta)={\delta\phi}_{\vec k,i}\,f(k\eta)$,
where ${\delta\phi}_{\vec k,i}$ is the initial perturbation amplitude
and $f(k\eta)$ is a dispersion factor.
For a super-horizon mode with $k\eta\ll 1$, $f(k\eta)=1$; the factor then
oscillates with a decaying envelope once the mode enter the horizon.
Let us define the initial power spectrum $P_{\delta\phi}(k)$ by
$\left<\delta\phi_{\vec k,i}\delta\phi_{\vec k',i}\right> = (2\pi^2/k^3)
P_{\delta\phi}(k) \, \delta({\vec k}-{\vec k}')$. We solve for $f(k\eta)$ numerically
using Eq.~(\ref{fourierphi}) with $\dot{\bar\phi}=0$ and the initial power spectrum
$P_{\delta\phi} (k)=Ak^{n-1}$, where $A$ is a constant amplitude squared and $n$ is the spectral index.
The space-time background has no difference from that of the Lambda Cold Dark Matter (LCDM) model.
Assuming a scale-invariant spectrum ($n=1$) and a combined constant parameter $A\beta^2$,
the induced $B$-mode polarization is computed using our full Boltzmann code based on the CMBFast~\cite{SZ}.
We have tuned the values of $r$ and $A\beta^2$, by fixing the other cosmological parameters to the 
best-fit values of the {\em Planck} 6-parameter LCDM model~\cite{planck}, to best fit the WMAP, {\em Planck}, BICEP2, and POLARBEAR data as shown in Fig.~\ref{fig:BMODE}. 
The likelihood plot in Fig.~\ref{likelihood} shows the maximum likelihood values of $r=0.168$ and $A\beta^2=0.0046$.
We have also produced the rotation power spectrum~\cite{Li,Caldwell,LLN2},
\be
C_l^{\alpha}=\left<|\alpha_l^m|^2\right>=\frac{\beta^2}{2\pi}
\int {dk}{k^2} \left\{ {\delta\phi}_k (\eta_s)\, j_l[k(\eta_0-\eta_s)]\right\}^2\,,
\label{clalpha}
\ee
where the rotation angle $\alpha({\hat n})=\sum_{lm} \alpha_l^m Y_l^m ({\hat n})$,
$\eta_0$ is the present time, and $\eta_s$ denotes the time when the primary CMB polarization is generated
on the last scattering surface or the rescattering surface. The rotation power spectra for the recombination and
the reionization with $A\beta^2=0.0046$ are shown in Fig.~\ref{fig:alpha}.
Using this rotation power spectrum, the rotation-induced $B$-mode polarization can be approximated by
\be
C_l^{BB}={1\over\pi}\sum_{l_1,l_2} (2l_1+1)(2l_2+1) C_{l_1}^{EE}(\eta_s) C_{l_2}^{\alpha}
\left( \begin{array}{ccc} l &l_1 &l_2\\  2 &-2 & 0\end{array} \right)^2\, ,
\label{cbl}
\ee
where we have assumed a negligible primary $B$ mode and used Wigner 3-$j$ symbols.
Recently, using the rotation-angle quadratic estimator $\hat{\alpha}_l^m$ 
of the correlation between CMB temperature and $B$-mode polarization ($TB$ power spectrum), 
constraints on direction-dependent cosmological birefringence from WMAP
7-year data have been derived, with an upper limit on the quadrupole of a scale-invariant
rotation power spectrum (i.e., $C_l^{\alpha}\propto l^{-2}$), $C_2^{\alpha}<3.8\times 10^{-3}$~\cite{Gluscevic}.
Our quadrupole is within this limit. In fact, the limit should become weaker for
our case because our $C_l^{\alpha}$ scales as $l^{-2}$ for $l<100$ and $l^{-4}$ for $l>100$.

Note that both $C_l^{TB}$ and $C_l^{EB}$ power spectra vanish due to the fact that $\left<\delta\phi\right>=0$.
On the other hand, the conversion of $E$ to $B$ would diminish both $C_l^{TE}$ and $C_l^{EE}$. We have found that
the effect of the birefringence on CMB power spectra predominately comes from low-$l$ $C_l^{\alpha}$. As such, the observed
power spectra basically follow the shapes of the original power spectra and can be approximated by
$C_l^{BB,{\rm obs}}=C_l^{EE} \sin^2 (2{\bar\alpha})$,
$C_l^{TE,{\rm obs}}=C_l^{TE} \cos (2{\bar\alpha})$, and
$C_l^{EE,{\rm obs}}=C_l^{EE} \cos^2 (2{\bar\alpha})$, where $\bar\alpha$ is a root-mean-square rotation angle.
Recent observations have constrained that $|\bar\alpha|< 3^\circ$~\cite{bicep1}. In the present consideration,
the birefringent $C_l^{BB}\sim 0.0025 C_l^{EE}$, implying a root-mean-square rotation angle of about $1.4^\circ$, which is
still within the above constraint. Certainly, a simultaneous fitting of these power spectra to all of the currently available polarization data
is required to give an accurate answer.

Recently, the gravitational lensing $B$-mode polarization has been detected
by cross correlating $B$ modes measured  by the SPTpol experiment with
lensing $B$ modes inferred from cosmic infrared background fluctuations measured by
Herschel and $E$ modes measured by SPTpol~\cite{spt}. Another CMB experiment called
POLARBEAR has also confirmed this cross correlation~\cite{polarbear}.
However, we note that this detection has no constraint on the rotation-induced $B$-mode polarization
because the rotation power spectrum and the lensing power spectrum are uncorrelated.

There have been physical constraints on $A$ and $\beta$. Let us assume that
inflation generates the initial condition for dark energy perturbation.
Then, $n\simeq 1$ and $A \simeq (H/2\pi)^2/M^2$, where $H$
is the Hubble scale of inflation. The recent CMB anisotropy measured by
the {\em Planck} mission has put an upper limit on $A< 3.4\times 10^{-11}$~\cite{planck}.
This implies that the present spectral energy density of
dark energy perturbation relative to the critical energy density,
$\Omega_{\delta\phi}< 10^{-15}$, which is negligible compared
to that of radiation. The most stringent limit on $\beta$ comes from
the absence of a $\gamma$-ray burst in coincidence with
Supernova 1987A neutrinos, which would have been converted in the galactic magnetic field
from a burst of axion-like particles due to the Primakoff production in the supernova
core: $\beta < 2.4\times 10^7$ for $m_\phi < 10^{-9}{\rm eV}$~\cite{sn1987}.
Hence the combined limit is $A\beta^2<2\times 10^4$, which is much bigger
than the value that we have used here.

The removal of dust contamination may reduce the tensor contribution to $r=0.16$~\cite{bicep2}. As such, the contours in Fig.~\ref{likelihood} are simply shifted vertically down by about a value of $0.04$. With nonzero $A\beta^2=0.005-0.009$, the solid 
$1$-sigma contour shows $r=0.09-0.1$, which are within the {\em Planck} upper limit of $r<0.11$. Thus, it might be too hasty to conclude that many inflation models with small $r$ are ruled out based on BICEP2 result. With the proper mechanisms like birefringence to induce the large-scale B-mode polarization, many inflation models can be still compatible with BICEP2 result, at least at 1-sigma level.

\begin{figure}[htbp]
\centerline{\psfig{file=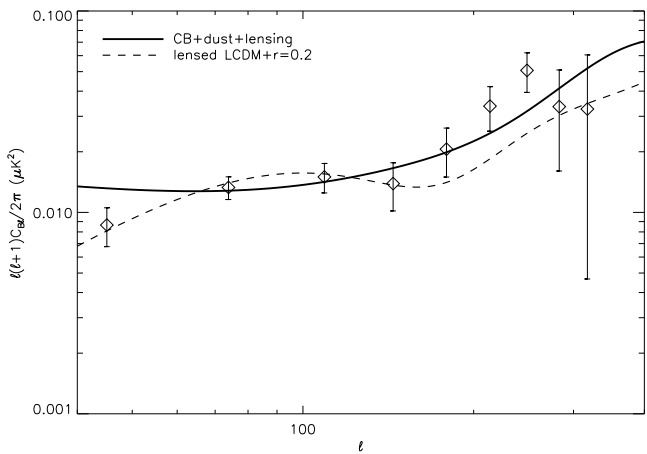, width=9cm}
} \caption{Solid curve is the $B$-mode power spectrum drawn with dust power $D=0.01$,  $A\beta^2=0.0023$, and $r=0$. Dashed curve is that from the BICEP2 fitting with $r=0.2$.}
\label{fig:dustCB}
\end{figure}

\begin{figure}[htbp]
\centerline{\psfig{file=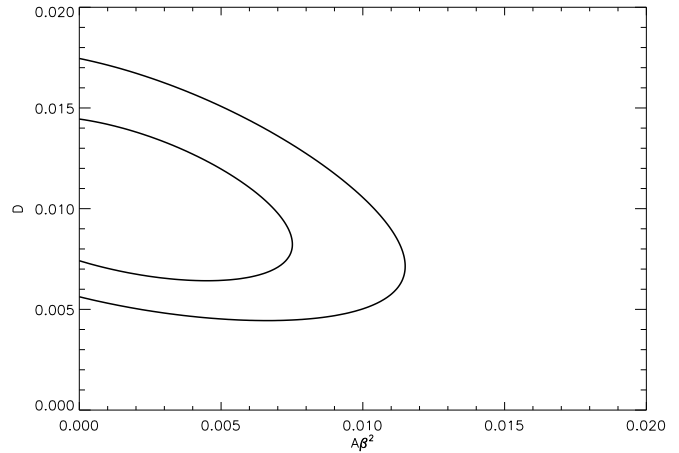, width=9cm}
} \caption{Likelihood plot of the parameters $D$ and $A\beta^2$, showing $1$-sigma and $2$-sigma contours. The maximum likelihood values are $D=0.01$ and $A\beta^2=0.0023$.}
\label{fig:dustCB_contour}
\end{figure}

\begin{figure}[htbp]
\centerline{\psfig{file=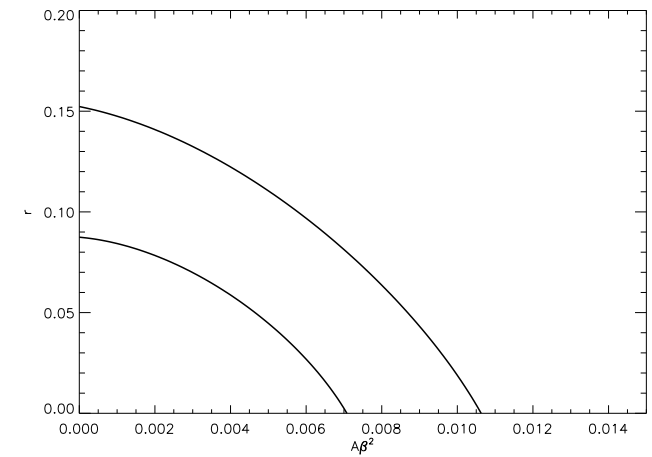, width=9cm}
} \caption{Likelihood plot of the parameters $r$ and $A\beta^2$, showing $1$-sigma and $2$-sigma contours.}
\label{fig:rCB_contour}
\end{figure}

Recently, a joint analysis of BICEP2/Keck Array and {\em Planck} data that includes dust contribution have yielded an upper limit $r<0.12$~\cite{b+p}. Here we have used a dust $B$-mode power spectrum, $l(l+1) C^{BB}_{lD}/(2\pi)=D (l/100)^{-0.3} {\rm\mu K^2}$ with a free parameter $D$, which was adopted in Ref.~\cite{mortonson}, combined with the birefringence $B$ modes and the tensor mode to fit the BICEP2 and POLARBEAR data. The best-fit power spectrum is drawn with $D=0.01$, $A\beta^2=0.0023$, and $r=0$, as shown in Fig.~\ref{fig:dustCB}. This power spectrum ( with $\chi^2=4.83$) is a better $\chi^2$ fitting than that of the pure tensor mode with $r=0.2$ ($\chi^2=7.69$). Fig.~\ref{fig:dustCB_contour} is the likelihood plot of the dust power and birefringence and Fig.~\ref{fig:rCB_contour} is that of the tensor and birefringence. The detection level of the dust power and the upper limit on $r<0.09$ are close to those found in Ref.~\cite{mortonson}, while we can set an upper limit on $A\beta^2<0.0075$.

Cosmological birefringence perturbation can generate a rotation-induced $B$-mode
power spectrum. The BICEP2 and POLARBEAR experiments may have barely $1$-sigma detection of 
cosmological birefringence $B$ modes at degree and sub-degree angular scales, if dust emission is ignored.
When dust foreground is included, we have set an upper limit on the cosmic birefringence.
It would be very important to make precise direct measurements of
$B$-mode polarization at sub-degree scales where birefringence $B$ modes peak at and can be mixed with lensing $B$ modes.
It thus poses a big challenge to do the separation of different $B$-mode signals. It is apparent that the rotation-induced
$B$-mode has acoustic oscillations but to detect or rule out them will require next-generation experiments.
In principle, one may use de-lensing methods~\cite{delensing}
or lensing contributions to CMB bi-spectra~\cite{bispectra} to single out the lensing $B$ modes.
Furthermore, de-rotation techniques can be used to remove the rotation-induced
$B$ modes~\cite{kamion09}. More investigations along this line should be done before
we can confirm the detection of the genuine $B$ modes.

This work was supported in part by the National Science Council, Taiwan, ROC under the Grants
No. NSC101-2112-M-001-010-MY3 (K.W.N.) and NSC100-2112-M-032-001-MY3
(G.C.L.).

\end{document}